# Ultrasonic Blind Stick For Completely Blind People To Avoid Any Kind Of Obstacles

Arnesh Sen  Kaustav Sen  Jayoti Das

Jadavpur University: Dept. of Physics, Kolkata, India

senarnesh.elec@gmail.com  sen.kaustavphys@gmail.com  jayoti.das@gmail.com

*Abstract*—The ability to live without being controlled by any action, judgment and any outside factors including any opinions and regulations is defined by the term Independent. But in reality physical movement for travelling or simply walking through a crowded street pose great challenge for a visually impaired person. Also they must learn every detail about the home environment such as placement of tables; chairs etc. to prevent injury. Because of this disability they have to sacrifice their independence in daily living by depending on the sighted people in every busy place like bus, footpaths, railway stations etc. This paper aims to design an artificial navigating system with adjustable sensitivity with the help of ultrasonic proximity sensor to assist these blind persons to walk fearlessly and independently in both indoor and outdoor environment. This system can detect any type of upcoming obstacles and potholes using the reflection properties of ultrasound. Attachment of the system to different body areas makes its utilization more versatile and reliable.

## I. INTRODUCTION

The purpose of this project is to get rid of those excuses caused by lack of ability to take complete control of movement for visually disabled people through a crowded place along with enormous types of obstacles which is the well-known characteristics of busy city life. Every day during competing in the global race to reach absolute perfectness, lifestyle of people completely depends upon technology which result Eye Dog and White Cane to compensate this ability-gap somehow. But in the widen horizon these technical aids are proven not fully reliable as they failed to overcome the super crazy situations, arises in modern streets especially in busy market places. Our society still contains about 285 million visually impaired, 39 million blind and 246 poor vision infested people as confirmation of WHO [1]. Well, though it is a myth to say completely good bye to this ability-gap, but there are still some innovative modifications are exposed in this paper which can be applied to existing technology to minimize this gap.

Reflection properties of ultrasound [2] is the prime key of all artificial navigation systems since then it was set forth with Lazzaro Spallanzani by observing the *bat echolocation* which also results a eeureka moment when the same logic is applied to find Titanic in underwater by Lewis Richardson in the year 1912. Brain Port [3], Ultra cane [4], Ray device [5], Eye Dog [6], White Cane [7] are all the torch bearer of the generation flux of that innovation. The effort of this paper is to make this journey smooth by implementing ultrasonic modules more efficiently to cover most of the surroundings including the detection of upcoming potholes as additional facility.

## II. DETECTION OF HEAD CHEST AND WAIST LEVEL OBSTACLES

Before starting it would be strongly recommended to study Table-IV as it contains the definitions all of the technical terms used in this paper to explain the working principle with great details.

### A. Differenciate between Head Chest and Waist Level

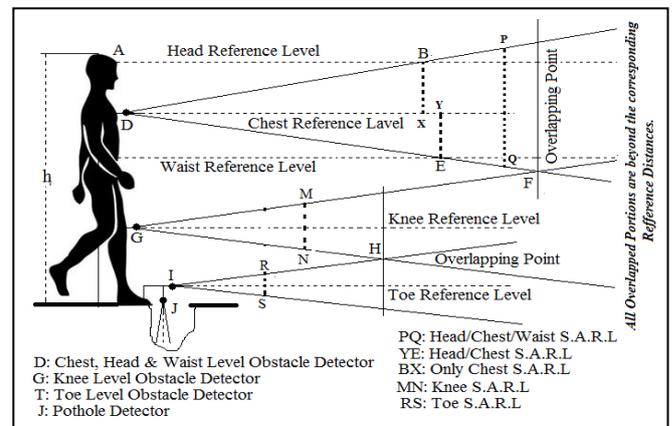

D: Chest, Head & Waist Level Obstacle Detector
G: Knee Level Obstacle Detector
T: Toe Level Obstacle Detector
J: Pothole Detector

PQ: Head/Chest/Waist S.A.R.L
YE: Head/Chest S.A.R.L
BX: Only Chest S.A.R.L
MN: Knee S.A.R.L
RS: Toe S.A.R.L

Figure-1 : All Sensors are activated together

The self-explanatory term chest level sensor [5] is used to detect the obstacles approaching towards the user in chest level and as well as the diverging property of the module which is 30 degree is allowed efficiently to handle the head level as well as waist level obstacles also by setting the solid angle reference line (S.A.R.L) in different position (See Table-I) [8]. If any kind of upper level type obstacles appear somewhere in around 150 cm far from the user chest level sensor corresponding buzzer (BrzC) will activate in Level-1and user should move forward carefully. When this distance is between 87 to 150 cm and if BrzC will suddenly deactivate then user should stop for a while and should slightly move back to check BrzC will activate again (in Level-2) or not. If BrzC will activae again then it confirms to the user that the obstacle is definitely around waist level. As waist level S.A.R.L will face obstacle before head level and chest level S.A.R.L (See Figure-1). If similar situation occur between 60 to 87 cm and in between 40 to 60 cm user can conclude the position of obstacle in head and chest level respectively. User can differentiate the range of distances by the same buzzer with different sound level (Level-1, 2, 3 and 4 respectively).

TABLE-I.    AVOIDING OF UPPER LEVEL OBSTACLES

| Position of Sensors | Sensing Range in cm | Possible Sensing Areas | Buzzers Action if Obstacles are Detected | | Position of Overlapping Point with Adjacent Sensor from Source | |
|---|---|---|---|---|---|---|
| | | | Buzzer's Name | Alarming Level | Name | Distance (cm) |
| Chest | 0 to 150 | Head/Chest/Waist | BrzC | 1 | Chest to Knee | 187 |
| | 0 to 87 | Head/Chest | BrzC | 2 | | |
| | 0 to 60 | Head | BrzC | 3 | | |
| | 0 to 40 | Chest | BrzC | 4 | | |
| Knee | 0 to 60 | Knee & Surroundings | BrzK | 1 | Knee to Toe | 84 |
| | 0 to 30 | | BrzK | 2 | | |
| | 0 to 10 | | BrzK | 3 | | |
| Toe | 0 to 40 | Toe & Surroundings | BrzT | 1 | Toe to Arch | None |
| | 0 to 20 | | BrzT | 2 | | |
| | 0 to 10 | | BrzT | 3 | | |

### B. Position of the Detector and Overlapping Problem

The sensor (chest level) is attached accordingly and there is a question of overlapping occurs to the knee level sensor (See Figure-1). As the divergence angle is 30 degree for both the sensors simple trigonometric calculation result the distance of overlapping point from the user of height 175 cm [9] is 187 cm and the position farthest position of S.A.R.L with respect to chest level sensor is set at 150 cm. Thus the working principle of the system is unaffected as HCSR04 is programmed to activate maximum up to S.A.R.L point.

### III. DETECTION OF LOW LEVEL OBSTACLES AND STAIRCASES

#### A. Lower Level Obstacle Detection

Logic used to determine lower level obstacles of any type (See Table I) is exactly same as the detection of upper level obstacles with a different position of S.A.R.L.

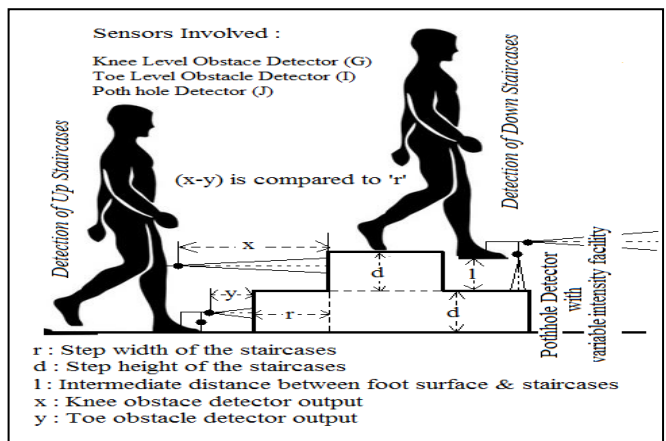

Figure 2 : Detection of Up and Down Stairs

TABLE-II.    AVOIDING OF LOWER LEVEL OBSTACLES

| Sensors Involved | Position of Solid angle boundary (cm) | Ultrasound Reflection Occurred | 24cm <(x-y) <26cm | Action of Buzzers | | Reaction of User |
|---|---|---|---|---|---|---|
| | | | | BrzK (Level) | BrzT (Level) | |
| Knee | 40 | Yes | Yes | 1 | 1 | Assume up Stairs ahead |
| Toe | 20 | Yes | | | | |
| Knee | 40 | Yes | No need to Check | 1 | 0 | Only Knee level obstacle ahead |
| Toe | 20 | No | | | | |
| Knee | 40 | No | No need to Check | 0 | 1 | Only Toe level obstacle ahead |
| Toe | 20 | Yes | | | | |
| Knee | 40 | No | No need to Check | 0 | 0 | Move Forward without hesitation |
| Toe | 20 | No | | | | |

As previous here is a also chance of overlapping between knee and toe level sensor but the same trigonometric calculation reveals that overlapping occurs at the distance of 87 cm from the user and farthest position of S.A.R.L is programmed at 60 cm that means user will not be confused due to overlapping (See Table). There is no chance of overlapping between toe level sensor and pot-hole detector as pothole detector is oriented vertically downward and emits ultrasound with same divergence angle 30 degree.

#### B. Detection of Up-Stairs

The co-ordination of knee level and toe level sensor is efficiently used to detect up staircases. The difference (x-y) of the estimated distances, measured by these two sensors is calculated. If the difference is sufficient to place the human foot with standard length (minimum 23 to 28 cm) it can be predicted as a stair step. Actually the standard foot length of a normal person is generally around 22 cm to 25 cm ( Kids have different lengths) , here the system is programmed to respond in a slightly higher value to avoid the chances of accidents . The response of corresponding buzzers is described in Table-. (See Table II).

#### C. Detection of Poth-Holes and Downstairs

The potholes detector sensor can use to detect both the potholes and the downstairs. This sensor is programmed in a logic which is exactly opposite to the program used to detect the front side obstacles. At first the distance between the sensor and the ground is detected. If this distance is within 0.15 to 0.30 m then it is considered as a down step [10] of a stair according to figure 3 and if the distance is larger than that range then it is considered as a pothole. Now according to the response of level of the corresponding buzzer user can also predict whether it is safe to move through the potholes or to avoid it (See Table III).

TABLE-III. DETECTION OF POTHOLES AND DOWNSTAIRS

| Sensor Position | Solid Angle Boundary Position | | Presence of Pot Hole | Depth of Poth Hole (in cm) | BrzP (Level) | Reaction of User |
|---|---|---|---|---|---|---|
| | Set in (cm) | Comment | | | | |
| Front Foot Arch | 10 | As foot arch in constantly swing in horizontal as well as vertical direction the detector will be programmed to accept minimum depth which is greater than 10 cm as pot hole. | No | 0 – 10 | 0 | Move Forward |
| | | | Yes | 10 – 20 | 1 | Move Forward with caution (It may be steps of downstairs) |
| | | | | 20 – 40 | 2 | Move Forward with Alternate Path |
| | | | | Beyond 40 | 3 | Stop Movement Immediately |

.

## IV. RELIABILITY AND REFRESH RATE OF THE SYSTEM

Here the reliability test of the system is analyzed by assuring by this set of raw data and after calibrating these data plot of actual distance vs. measured distance of the sensor is shown [11]. The minimum and maximum distance measuring capacity of the sensor is 0.03 m to 3 m with a measuring angle of 30 degree. The refresh rate of HCSR04 is controlled by the programmer. Feasibility test reveals that the 30ms interval results best obstacle handling capacity [12].

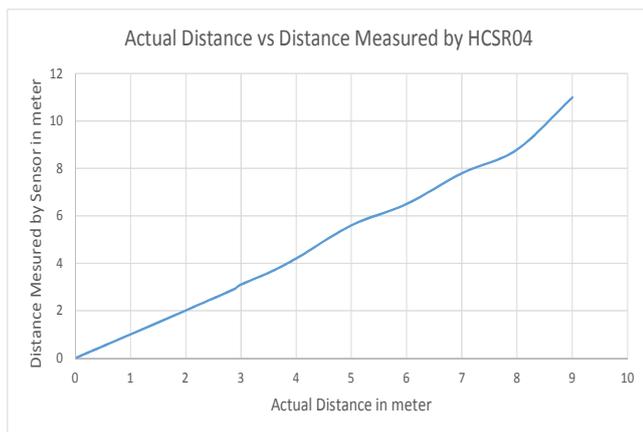

Figure 3 : Actual Distance versus Measured Distance by the Sensor

## V. NOVELTY AND SCOPE OF IMPROVEMENT

### A. Novelty of the System

The co-ordination of the conventional sensors and detectors make the navigation process unique and prepare to handle the variety of obstacle comes from various front direction. It is also able to detect potholes, up-stairs and down-stairs. The verticality of the system over the all existing method is undoubtedly reliable and exactly at this point this paper demands its novelty. Feasibility test reveals that the performance is outstanding both in indoor and outdoor environment with a greater safety.

### B. Limitations and Scope of Improvement

1) *Limitations of the System :*

a) Sensing accuracy can be affected by change of temperature of 10 degrees or more and can also be affected by soft matirials.

b) Minimum obstacle thickness sensing capcity 3mm and minimum obstacle distance should be 2cm for detection.

c) The user have to havituated to the use of this system before ause it in outdoor environment.

2) *Scope of Improvement :*

a) By attaching microservos to the detectors the divergence range can be incresed significantly.

b) By make the detectors able to differentiate between closely valued accoustic impedance of the mediums the system can be improved to detect the condition of the surface.(Whether the surface is contained with mud, stagnent water).

c) The buzzers can be replaced by artificial vibrators as most of the time user may not able to differentiate between the sound levels in noisy environment.

## VI. REFERNCE TABLE

TABLE IV. DEFINITIONS OF THE TYPICAL TERMS USED IN THE PAPER

| Term Used | | Definition |
|---|---|---|
| Solid Angle Reference Level (S.A.R.L) | | Vertical Reference plane Positioned in certain distance from the sensor. Sensor is programmed to activate when obstacles are located in this certain range |
| Details of Buzzers | BuzC | Corresponding Buzzer of Chest Level Sensor |
| | BuzK | Corresponding Buzzer of Knee Level Sensor |
| | BuzT | Corresponding Buzzer of Toe Level Sensor |
| | BuzP | Corresponding Buzzer of Pothole Detector |
| Level of Buzzers | 0 | Buzzer if Off |
| | 1 to 4 | Proportional to the sound level |

## VII. CONCLUTION

This upgraded navigation system is tasted more and more time in the all mentioned critical situations to claim the certificate of excellence in feasibility test. It is strongly expected that with more necessary modification the movement ability gap can be made more minimize in near future.

## VIII. ACKNOWLEDGEMENT

We would like to show gratitude to Dr. Jayoti Das, Associate Professor, Dept. of Physics, Jadavpur University for her kind support, constant supervision and encouragement during this course of research work which will help us to reach intended goal in the midst of intermingled maze of possibilities and failures.

## References

[1] World Health Organization. Visual impairment and blindness, 2011. Available: http://www.who.int/mediacentre/factsheets/fs282/en/

[2] Denis Tudor, Lidia Dobrescu, Drago? Dobrescu, "Ultrasonic electronic system for blind people navigation," 5th IEEE International Conference on E-Health and Bioengineering - EHB 2015 Grigore T. Popa University of Medicine and Pharmacy, Ia?i, Romania, November 19-21, 2015.

[3] BrainPort Technologies. BrainPort® V100, 2015. Available http://www.wicab.com/en_us/v100.html

[4] Ultracane, 2015. Available http://www.ultracane.com/

[5] Sanchez, J.; Oyarzun, C. Mobile Assistance Based on Audio for Blind People Using Bus Services. In New Ideas in Computer Science Education (In Spanish); Lom Ediciones: Santiago, Chile, 2007; pp. 377–396.

[6] Mahdi Safaa A., Muhsin Asaad H. and Al-Mosawi Ali I. Using Ultrasonic Sensor for Blind and Deaf persons Combines Voice

[7] Regalado Melo Rosa P., Bengala de apoio a cegos com deteção de buracos, Dissertação de Mestrado, Universidade de Aveiro, 2009.

[8] Pinedo, M.A.; Villanueva, F.J.; Santofimia, M.J.; López, J.C. Multimodal Positioning Support for Ambient Intelligence. In Proceedings

[9] Ing. Jaroslav ILEČKO, "The simulation of human gait in Solid Works," Department of production systems and robotics, Faculty of Mechanical Engineering, Technical University of Košice, unpublished.

[10] Bonnie Yuk San Tsung, Ming Zhang, Yu Bo Fan, David Alan Boone, "Quantitative comparison of plantar foot shapes under different weight-bearing conditions," Volume 40 Number 6, November/December 2003 Pages 517 — 526.

[11] Hub, A.; Hartter, T.; Ertl, T. Interactive Localization and Recognition of Objects for the Blind. In Proceedings of the 21st Annual Conference on Technology and Persons with Disabilities, Los Angeles, CA, USA, 22–25 March 2006; pp. 1–4.

[12] Sonnenblick, Y. An Indoor Navigation System for Blind Individuals. In Proceedings of the 13th Annual Conference on Technology andPersons with Disabilities, Los Angeles, CA, USA, 17–21 March 1998; pp. 215–224.